\documentclass[twocolumn,smallcondensed,smallextended]{svjour2}
\smartqed  
\usepackage{graphicx}
\usepackage{amssymb}
\usepackage{color}
\newtheorem{Th}{Theorem}
\newtheorem{Cor}{Corollary}
\newtheorem{Con}{Conjecture}
\newtheorem{df}{Definition}
\newtheorem{Rem}{Remark}

\begin{document}

\title{Generalized Arcsine Law and Stable Law in an 
Infinite Measure Dynamical System
}


\author{Takuma Akimoto         
}


\institute{T. Akimoto \at
              Department of Applied Physics, Advanced School of Science and Engineering, Waseda University, Okubo 3-4-1, Shinjuku-ku, Tokyo 169-8555, Japan. \\
              Tel.: +81-3-3200-2457\\
              Fax: +81-3-3200-2457\\
              \email{akimoto@aoni.waseda.jp}           
}

\date{Received: date / Accepted: date}

\maketitle

\begin{abstract}
Limit theorems for the time average of some observation functions in 
an infinite measure dynamical system
 are studied. It is known that
 intermittent phenomena, such as the Rayleigh-Benard convection and 
Belousov-Zhabotinsky reaction, are described by infinite measure
 dynamical systems.
We show that the time average of the observation function 
which is not the $L^1(m)$ function, whose average 
with respect to the invariant measure $m$ is finite, 
converges to
the generalized arcsine distribution. 
This result leads to the novel view
 that 
the correlation function is intrinsically random and does not
decay. Moreover, it is also numerically 
shown that the time average of the 
observation function converges to the stable distribution 
when the observation function has the infinite mean.

\end{abstract}

\section{Introduction}
\label{intro}

Recently, the $1/f$ power spectrum and the power law phenomena, which 
are closely related to the intermittent phenomena \cite{M}, have 
been studied
 in various systems. Examples are the $1/f$ power spectrum in 
the Rayleigh-Benard convection \cite{Ah}, 
Belousov-Zhabotinsky reaction \cite{P}, fluorescence intermittency in single 
nanocrystals \cite{Fl}
and the power law decay of the earthquake phenomena \cite{Om,Bott}. 
It is also known that 
such power law phenomena are clearly observed in Hamiltonian systems 
and non-hyperbolic dynamical systems \cite{Ai1,Ai6}. It is a remarkable 
problem that
 the theoretical 
meaning of the $1/f$ power spectrum has not been completely elucidated.\par
In ergodic theory the time average can be replaced 
by the ensemble average. However, in the intermittent phenomena, this replacement is not always guaranteed. Actually,
 it has been pointed out that the time average of 
some observation shows anomalous behaviour, that is, the time average does not 
converge to a constant value and becomes intrinsically random \cite{Brok,Marg}.
Therefore it is important to analyze 
the behaviour of the time average in the intermittent phenomena from the 
viewpoint of ergodic theory.
Infinite measure dynamical systems are examples of dynamical systems 
 describing such intermittent phenomena. The recent study of
infinite ergodic theory tells us 
 that the time average of some observation
 functions converges in distribution \cite{DK,Aa,T,TZ}. 
For example, the scaled time average of the $L^1(m)$ function $g(x)$, $\sum_{k=0}^{n-1}g(T^k\cdot)/a_n$, 
converges to the Mittag-Leffler distribution, where $a_n$ is the proper sequence.\par
 The purpose of this paper
 is to study the time average of the non-$L^1(m)$ function in 
infinite measure systems. In the previous works, it is already 
known that the distribution 
of the occupation time of some interval in infinite measure systems converges to the generalized  arcsine distribution \cite{T4,Bel}.
In this paper
we clear the class of the observation function whose time average converges to 
the generalized arcsine distribution using the modified Bernoulli map which is the typical example of 
a one-dimensional infinite measure dynamical system on $[0,1]$.
\par
The paper is organized as follows. 
In \S 2 we review the modified Bernoulli map. In \S 3 we show the distribution 
for the time average of the $L^1_{loc}(0,1)$ function with finite mean  
using a modified Bernoulli map. Theorem 1 and Theorem 2 are the 
generalized arcsine laws, which are well known in random process, in the dynamical system. Theorem 3, which is the main result in this paper, generalizes the observation function.  In \S 4 our results are applied to the 
correlation functions. In \S 5 we numerically demonstrate the distribution of 
for the time average of the $L^1_{loc}(0,1)$ function with infinite mean.
\S 6 is devoted to summary and 
an approach toward the ergodic problems
 of non-equilibrium statistical mechanics.

\section{Reviews of the modified Bernoulli map}
\label{sec:1}
In this section we review the statistical properties of the modified Bernoulli 
map. The modified Bernoulli map is one dimensional map on $[0,1]$ defined 
by
\begin{eqnarray}
x_{n+1} &=& Tx_n \nonumber\\
&=&\left\{
\begin{array}{ll}
 x_n + 2^{B-1}x_n^B &
\qquad x_n\in I_0=[0,1/2]\\
\noalign{\vskip0.2cm}
 x_n - 2^{B-1}(1-x_n)^B &
\qquad x_n\in I_1=(1/2,1].
\end{array} \right.
\label{eq:2.1}
\end{eqnarray}
This map has two indifferent fixed points ($x=0$ and $1$), at which the 
invariant density $\rho (x)$ is not bounded. The invariant density $\rho (x)$
can be written as 
\begin{equation}
\rho(x) \sim x^{1-B}+(1-x)^{1-B}
\label{eq:2.2}
\end{equation}
for $x\sim 0$ and $x\sim 1$.
Therefore
for $B\geq 2$ the invariant density cannot be normalized, that is, the invariant 
measure becomes the infinite one \cite{Ai1,Ai2,T2}.\par
It is important that renewal processes are constructed by the sequences of the 
modified Bernoulli map \cite{Cox}. Actually, using the symbolic sequence 
$\sigma_n=\sigma (x_n)$, where $\sigma (x)=-1$ $(x\in I_0)$ and 
$\sigma (x)=1$ $(x\in I_1)$, one can define the renewal when the value of 
$\sigma_n$ changes, namely $\sigma_n\sigma_{n+1}=-1$. As
the time interval between $r-1$th and $r$th renewal denoted by ${\bf X}_r$ 
$(r\geq 2)$
is an independently identically distributed random variable, and this
 probability
 density function (p.d.f.) 
 $f(n)$ is given by
$f(n) \propto (n-1)^{-\beta}~(\beta=B/(B-1))$ 
for $n\gg 1$ \cite{Ai3}, this process is regarded as a renewal process.
 In what follows, we consider the time interval 
between successive renewals as continuous random variables, i.e., p.d.f. is 
given by 
\begin{equation}
f(x) = (\beta-1) (x-1)^{-\beta}~~~(\beta=B/(B-1)).
\label{eq:2.3}
\end{equation} 
It is noted that the p.d.f. of the first renewal time ${\bf X}_1$ depends on 
the initial ensemble of the modified Bernoulli map \cite{Ak1,Ak2}. When
the initial ensemble is the invariant density for the first passage map 
$T^{n(x)}(x)$ with respect to 
$E=[e_1,e_2]$, whose endpoints are the solutions of the equation 
$Tx=1/2$ for $e_1<e_2$ and 
$n(x)=1+\min \{n\geq 0:T^{n}(x)\in E\}$, the p.d.f. of ${\bf X}_1$ 
is same as Eq. (\ref{eq:2.3}), namely the ordinary renewal process.  
However,
when the initial ensemble is the invariant density for the modified 
Bernoulli map, the p.d.f. of the first renewal time ${\bf X}_1$ 
is given by $f_1(x)=(1-F(x))/\mu$, which
is completely different from Eq. (\ref{eq:2.3}), where 
$F(x)$ is the cumulative distribution function of $f(x)$ and $\mu$ 
is the mean value of ${\bf X}_r$ \cite{Cox}. In the previous papers 
\cite{Ak1,Ak2} we clearly demonstrate the dependence of the 
statistical laws, namely, the renewal function and the correlation 
function, on the initial ensemble.

\section{Generalized arcsine law}
Firstly, we analyze the behaviour of the time average of the following 
function:
\begin{equation}
I(x)=\left\{
\begin{array}{ll}
a &(x\leq \frac{1}{2})\\
b &(x>\frac{1}{2}),
\end{array}
\right.
\label{eq:3.1}
\end{equation} 
where $a,b\in \mathbb{R}\setminus \{-\infty,\infty\}$.\footnote{The special case, namely $a=-1$ and $b=1$, is shown by using the renewal theory in \cite{Ai5}.} It is noted that $I(x)$ is not the $L^1(m)$ function.
\par
We review Lamperti's generalized arcsine law 
for the modified Bernoulli map \cite{Lam}.\par

\begin{Th}[Lamperti's generalized arcsine law]
Let ${\bf X}_n$ be the time interval between the successive renewals in the 
renewal process constructed by the modified Bernoulli map and $N_n$ be the 
occupation time in $I_0$, that is, $N_n={\bf X}_1+{\bf X}_3+...+{\bf X}_n$ 
when $x_0\in I_0$ and $n$ is odd. Then 
\begin{equation}
\mathop{\lim}_{n\rightarrow\infty}\Pr(N_n/n\leq x)=G_{\alpha}(x)
\label{eq:3.2.1}
\end{equation} 
exists, where $\alpha=\beta-1$ and the p.d.f. $G'_{\alpha}(x)$ is given by
\begin{equation}
G_{\alpha}'(x)=
\frac{\sin \pi\alpha}{\pi}
\frac{x^{\alpha}(1-x)^{\alpha-1}+x^{\alpha-1}(1-x)^{\alpha}}{x^{2\alpha}
+2x^{\alpha}(1-x)^{\alpha}\cos\pi\alpha +(1-x)^{2\alpha}}.
\label{eq:3.2.2}
\end{equation} 
\label{Th:1}
\end{Th}

The distribution $G_{\alpha}(x)$ is called the generalized arcsine distribution.

{\it Proof}.
In \cite{Lam} the limit probability (\ref{eq:3.2.1}) exists if and only if 
there exist constants $c$ and $\alpha$ such that 
\begin{equation}
\mathop{\lim}_{n\rightarrow\infty} E(N_n/n)=c
\label{eq:3.2.2.1}
\end{equation}
 and 
\begin{equation}
\mathop{\lim}_{x\rightarrow 1-}\frac{(1-x)u'(x)}{1-u(x)}=\alpha,
\label{eq:3.2.2.2}
\end{equation}
 where $E(.)$ is the ensemble average with respect to the 
initial density and $u(x)=\sum_{n=1}^{\infty}f(n)x^n$ is the generating 
function of the $f(n)$. In what follows, we check the above conditions.
\par

First, we define the average of the function $1_{[0,1/2)}(T^nx)$ as
\begin{equation}
E_n\equiv E(1_{[0,1/2)}(T^nx)).
\label{eq:3.2.3}
\end{equation}
According to \cite{T3} the initial density $\rho_0(x)$ converges to the invariant density under 
proper normalization:
\begin{equation}
w_n P^n\rho_0(x)
\rightarrow \left(\frac{1}{\Gamma(\alpha)\Gamma(2-\alpha)}\right)
\rho(x)~~{\rm as}~n\rightarrow \infty,
\label{eq:3.2.4}
\end{equation}
where $P$ is the Perron-Frobenius operator and 
\begin{equation}
w_n \sim \left\{
\begin{array}{ll}
\log n  ~~&(B=2)\\
n^{1-\alpha}  &(B>2),
\end{array}
\right.
\label{eq:3.2.5}
\end{equation}
where $\alpha=\beta-1$. 
The invariant density $\rho(x)$ is symmetric with
 respect to the axis $x=1/2$. Therefore
\begin{equation}
E_n \rightarrow \frac{1}{2} ~~{\rm as}~n\rightarrow \infty,
\label{eq:3.2.5}
\end{equation}
and
\begin{equation}
E(N_n/n)=\frac{1}{n}\sum_{k=1}^{n}E_k\rightarrow \frac{1}{2}
~~{\rm as}~n\rightarrow \infty.
\label{eq:3.2.6}
\end{equation}

By Karamata's Tauberian theorem
it is easily confirmed that the condition (\ref{eq:3.2.2.2}) holds when 
$\alpha=\beta-1$:  
\begin{equation}
\mathop{\lim}_{x\rightarrow 1-}
\frac{(1-x)F'(x)}{1-F(x)}=\beta-1.
\end{equation}

The p.d.f. of $N_n/n$ in \cite{Lam} under
 $c=1/2$ and $\alpha=\beta-1$ 
implies the p.d.f. (\ref{eq:3.2.2}).
$\qed$

Using Theorem \ref{Th:1}, one can know the distribution of the time average of 
$I(x)$ immediately.

\begin{Th}
The time average of $I(x)$ converges in distribution:
\begin{equation}
\frac{1}{n}\mathop{\sum}_{k=0}^{n-1}I(T^k\cdot) \rightarrow
Y_{\alpha,a,b}
\label{eq:3.2}
\end{equation}
where the random variable $Y_{\alpha,a,b}$ has the following p.d.f,
\begin{eqnarray}
&&G_{\alpha,a,b}'(x)=\nonumber\\
&&
\left\{
\begin{array}{ll}
\frac{(a-b)\sin \pi\alpha}{\pi}
\frac{(x-b)^{\alpha-1}(a-x)^{\alpha-1}}{(x-b)^{2\alpha}
+2(x-b)^{\alpha}(a-x)^{\alpha}\cos\pi\alpha +(a-x)^{2\alpha}}
&(a>b)\\
\\
\frac{(b-a)\sin \pi\alpha}{\pi}
\frac{(b-x)^{\alpha-1}(x-a)^{\alpha-1}}{(b-x)^{2\alpha}
+2(b-x)^{\alpha}(x-a)^{\alpha}\cos\pi\alpha +(x-a)^{2\alpha}}
&(a<b),
\end{array}
\right.
\label{eq:3.3}
\end{eqnarray}
where $\alpha=\beta-1$, that is,
the random variables $Y_{\alpha,a,b}$ obeys the generalized arcsine law.
\label{Th:2}
\end{Th}

{\it Proof}. The time average of $I(x)$ can be rewritten as 
\begin{equation}
\frac{1}{n}\sum_{k=0}^{n-1}I(T^kx)=\frac{aN_n+b(n-N_n)}{n}.
\label{eq:3.4}
\end{equation}
Using Theorem \ref{Th:1}, we can write
\begin{eqnarray}
&&\Pr\left\{\frac{1}{n}\sum_{k=0}^{n-1}I(T^kx)\leq x\right\}
=\Pr\left\{(a-b)\frac{N_n}{n}+b\leq x\right\}\nonumber\\
&\rightarrow&\left\{
\begin{array}{ll}
G_{\alpha}\left(\frac{x-b}{a-b}\right)~~~~~&(a>b)\\
1-G_{\alpha}\left(\frac{x-b}{a-b}\right)~~~~~&(a<b).
\end{array}
\right.
{\rm as}~n\rightarrow\infty
\label{eq:3.5}
\end{eqnarray}
The derivative of (\ref{eq:3.5}), which is the p.d.f. of the time average of 
$I(x)$ denoted by $G_{\alpha,a,b}$, 
gives (\ref{eq:3.3}).\qed

\begin{df}[$L^1_{loc,m}$ function with finite mean]
If
the following conditions
\begin{equation}
\mathop{\lim}_{\epsilon\rightarrow 0}
\frac{\int_{\epsilon}^{1-\epsilon} |g|dm}
{\int_{\epsilon}^{1-\epsilon} dm}<\infty
\label{eq:3.6}
\end{equation}
and for all $\epsilon>0$
\begin{equation}
\int_{\epsilon}^{1-\epsilon} |g|dm<\infty
\label{eq:3.7}
\end{equation}
hold,
 then the function $g$ 
is called the $L^1_{loc}(0,1)$ function with finite mean 
with respect to $m$, denoted by $L^1_{loc,m}(0,1)$ function with finite mean.
\end{df}

Examples of the $L^1_{loc}(0,1)$ function w.r.t.
the invariant measure of the modified Bernoulli map are 
$I(x)$ and $g(x)=x$.\footnote{Roughly speaking, we can say the 
$L^1_{loc,m}(0,1)$ function with finite mean is considered as 
the $L^{\infty}(0,1)$ function.}

\begin{Rem}
In the case of the modified Bernoulli map the measure of 
the sets $[0,\epsilon]$ 
and $[1-\epsilon,1]$ are not finite. Therefore 
we exclude these sets in Eqs. $(\ref{eq:3.6})$ and $(\ref{eq:3.7})$.
When the measure of the other sets are not finite, 
Equations $(\ref{eq:3.6})$ and $(\ref{eq:3.7})$ must be changed 
to exclude these sets.
\end{Rem}

\begin{Th}
Let $g(x)$ be the $L^1_{loc}(0,1)$ function with finite mean
w.r.t.
the invariant measure $m$ and $g(0)=a,g(1)=b$. Further, there exists $\delta$ 
such that $0<\delta<1$ and $g(x)$ is continuous in $[0,\delta]\cup [1-\delta,1]$. Then the time average of $g(x)$ converges 
in distribution to $Y_{\alpha,a,b}$, where $\alpha=\beta-1$.\footnote{
The following proof is not changed even if the condition of $g(x)$ changes 
from the
 $L^1_{loc,m}(0,1)$ function to the $L^{\infty}(0,1)$ 
function.}
\label{Th:3}
\end{Th}

{\it Proof}. 
It is shown that the time average of the $L_+^1(m)$
function\footnote{
The $L_+^1(m)$ function is the $L^1(m)$ function 
whose value is positive.} 
converges to zero
\cite{Aa}, 
that is, for all $f\in L_+^1(m)$ 
\begin{equation}
\frac{1}{n}\sum_{k=0}^{n-1}f(T^kx)\rightarrow 0 \quad{\rm as}~n\rightarrow 
\infty~{\rm for~almost~all}~x.
\label{eq:3.8}
\end{equation}
By the theorem (\ref{eq:3.8}),  for all
$\epsilon>0$ there exist $N_1$ such that for 
$n>N_1$
\begin{eqnarray}
&&\left|\frac{1}{n}\sum_{k=0}^{n-1}g_{\delta,a,b}(T^kx)-\frac{1}{n}
\sum_{k=0}^{n-1}I(T^kx)\right|\nonumber\\
&&<
\frac{1}{n}\sum_{k=0}^{n-1}
\left|g_{\delta,a,b}(T^kx)-I(T^kx)\right|\nonumber\\
&&<\epsilon/2,
\label{eq:3.10}
\end{eqnarray}
where
\begin{equation}
g_{\delta,a,b}(x)=\left\{
\begin{array}{ll}
a ~~~~~~&{\rm for}~x\in [0,\delta)\\
g(x) &{\rm for}~x\in [\delta,1-\delta]\\
b &{\rm for}~x\in (1-\delta,1].\\
\end{array}
\right.
\label{eq:3.9}
\end{equation}
Then by the continuity of $g(x)$ and (\ref{eq:3.8}),  for all  
$\epsilon>0$ there exist $\delta_*$ and $N_2$ such that for $n>N_2$
\begin{equation}
\left|\frac{1}{n}\sum_{k=0}^{n-1}g(T^kx)-\frac{1}{n}
\sum_{k=0}^{n-1}g_{\delta_*,a,b}(T^kx)\right|<\epsilon/2.
\label{eq:3.11}
\end{equation}

Therefore for all $\epsilon>0$ there exists $\delta_*$ and $N$ such that 
for $n>N$ 
\begin{equation}
\left|\frac{1}{n}\sum_{k=0}^{n-1}g(T^kx)-\frac{1}{n}
\sum_{k=0}^{n-1}I(T^kx)\right|<\epsilon.
\label{eq:3.12}
\end{equation}
By Theorem \ref{Th:2}, the time average of $g(x)$ converges 
in distribution to $Y_{\alpha,a,b}$. \qed

\begin{figure}[h]
\includegraphics[width=8cm]{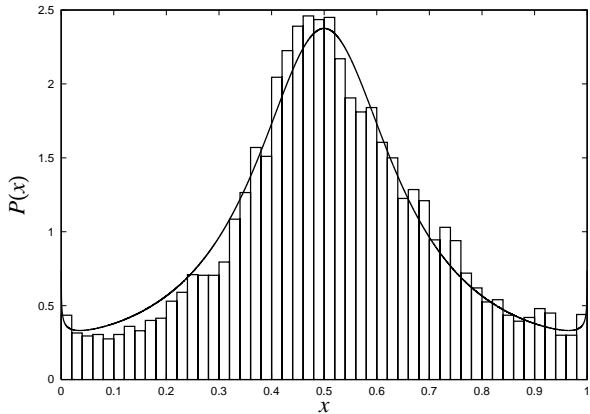}
\caption{The probability density function for the time average of 
$g(x)=x^2$ ($B=2.2$).
}
\label{fig1}
\end{figure}

\begin{figure}[h]
\includegraphics[width=8cm]{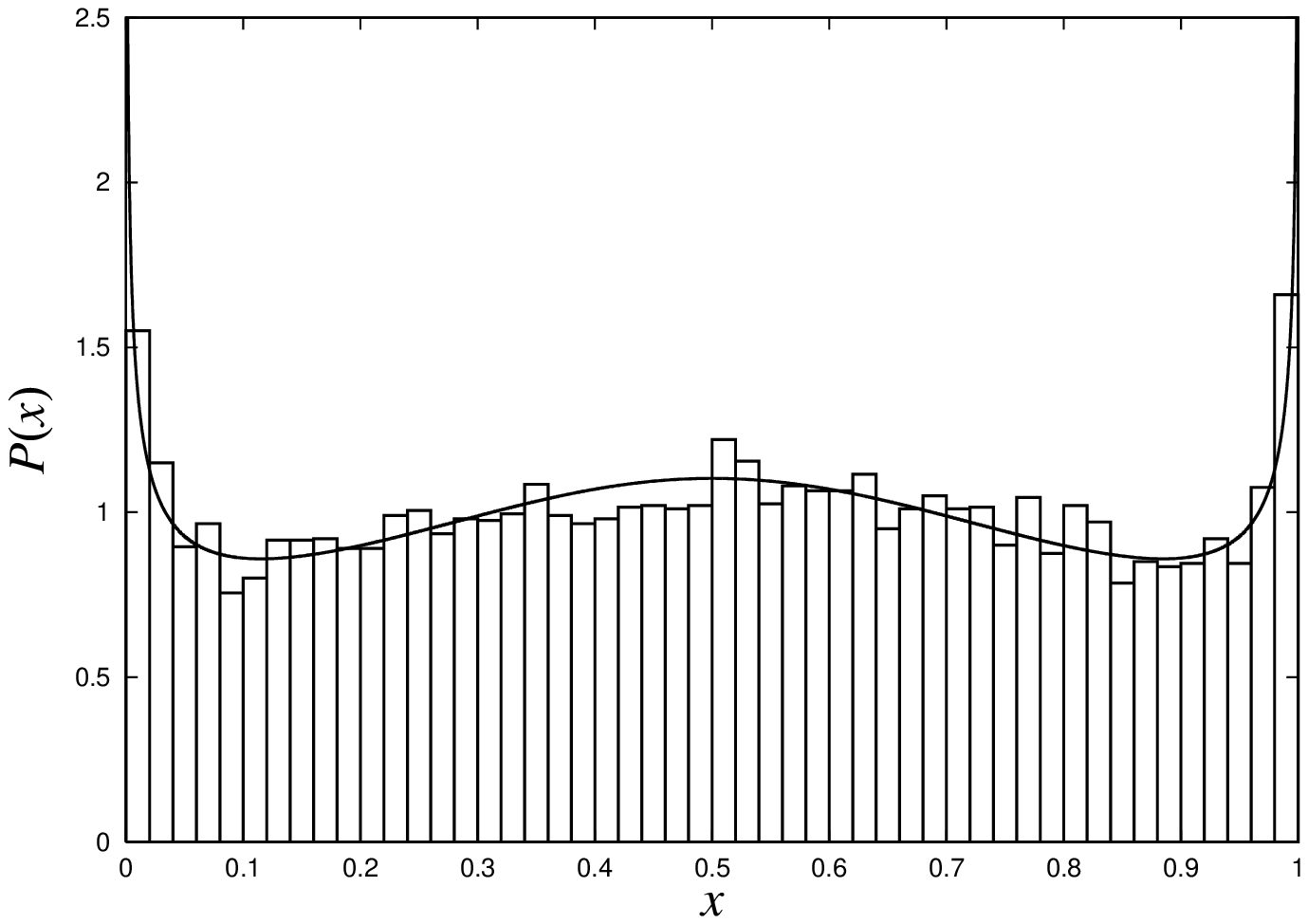}
\caption{The probability density function for the time average of 
$g(x)=x^2$ ($B=2.5$).
}
\label{fig2}
\end{figure}

\begin{figure}[h]
\includegraphics[width=8cm]{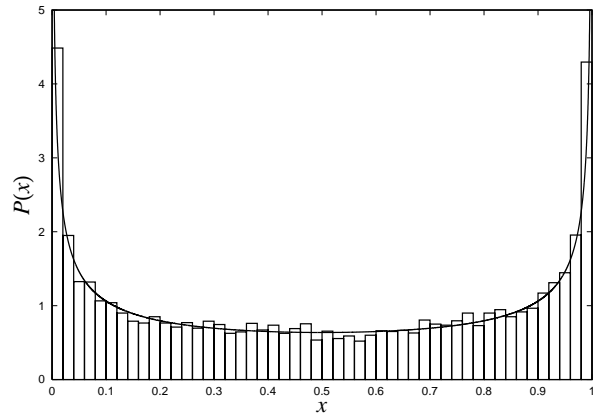}
\caption{The probability density function for the time average of 
$g(x)=x^2$ ($B=3.0$).
}
\label{fig5}
\end{figure}

Figures \ref{fig1}, \ref{fig2} and \ref{fig5} demonstrate numerically
that
the p.d.f. for 
the time average of $g(x)=x^2$ obeys the theoretical 
one even when the time $n$ is finite ($n=10^8$).

\section{Application to the correlation functions}

We apply Theorem \ref{Th:3} to the correlation function, which is defined by 
the time average:
\begin{equation}
C(n)=\mathop{\lim}_{N\rightarrow\infty}\frac{1}{N}\sum_{k=0}^{N-1}
g(x)h(T^nx).
\label{eq:4.1}
\end{equation}
In the case of finite measure, ergodic theory states that the correlation 
function defined by the time average equals to the correlation function 
defined by the ensemble average, namely the average of $g(x)h(T^nx)$ with 
respect to the invariant measure. However, when the invariant measure is not 
finite, the time-averaged correlation function is not equal to 
the ensemble-averaged correlation function.\footnote{The dependence of the 
decay of the ensemble-averaged correlation function on the initial ensemble 
is discussed in \cite{Ak2}.} In this section we demonstrate that 
the correlation function is intrinsically random in the modified Bernoulli 
map.\par

\begin{Cor}
For all $n$ the correlation function of $\sigma(x)$ converges to 1:
\begin{equation}
C(n)=\mathop{\lim}_{N\rightarrow\infty}\frac{1}{N}\sum_{k=0}^{N-1}
\sigma(x_k)\sigma(x_{k+n})=1,
\label{eq:4.2}
\end{equation}
where $\sigma(x)=1 ~(x\in [0,1/2)),$ $-1 ~(x\in [1/2,1])$.
\end{Cor}
{\it Proof}. 
For all $n$ the observation function $g_n(x)$ is defined as
\begin{eqnarray}
g_n(x)&=&\sigma(x)\sigma(T^nx)\nonumber\\
&=&\left\{
\begin{array}{ll}
+1\qquad &x\in [0,a_n] \cup [1-a_n,1]\cup A_n\\
-1 &{\rm otherwise}
\end{array}
,\right.
\label{eq:4.4}
\end{eqnarray}
where $a_{n}=a_{n+1}+2^{B-1}a_{n+1}^B$ ($a_n>0$ and $a_0=1/2$) and 
$A_n$ is the set which attains $\sigma(x)\sigma(T^nx)=1$ and is 
subset of $[a_n,1-a_n]$:
\begin{equation}
A_n=\{x\in [a_n,1-a_n] : \sigma(x)\sigma(T^nx)=1\}.
\label{eq:4.5}
\end{equation}
Then $g_n(0)=g_n(1)=1$ and $g_n(x)$ is continuous in $[0,a_n]\cup
[1-a_n,1]$
and for all $\epsilon>0$
\begin{equation}
\int_{\epsilon}^{1-\epsilon}g_n(x)dm<\int_{\epsilon}^{1-\epsilon}dm
<\infty.
\label{eq:4.8}
\end{equation}
By Theorem \ref{Th:3}, $C(n)$ is convergence in distribution to 
$Y_{\alpha,1,1}$ for 
all $n$.\qed

\begin{Cor}
For all $n$ the correlation function of $x$ is convergence in distribution to 
$Y_{\alpha,0,1}$:
\begin{equation}
C(n)=\mathop{\lim}_{N\rightarrow\infty}\frac{1}{N}\sum_{k=0}^{N-1}
x_kx_{k+n}\rightarrow Y_{\alpha,0,1}.
\label{eq:4.3}
\end{equation}
\end{Cor}

{\it Proof}. 
For all $n$ the observation function $g_n(x)$ is defined as
\begin{equation}
g_n(x)=x(T^nx),
\label{eq:4.6}
\end{equation}
and for all $\epsilon>0$
\begin{equation}
\int_{\epsilon}^{1-\epsilon}g_n(x)dm<\int_{\epsilon}^{1-\epsilon}dm
<\infty.
\label{eq:4.7}
\end{equation}
Then $g_n(0)=0$ and $g_n(1)=1$, and $g_n(x)$ is continuous in $[0,a_n]\cup
[1-a_n,1]$. 
By Theorem \ref{Th:3}, $C(n)$ is convergence in distribution to 
$Y_{\alpha,0,1}$ for 
all $n$.\qed

\begin{figure}[h]
\includegraphics[width=8cm]{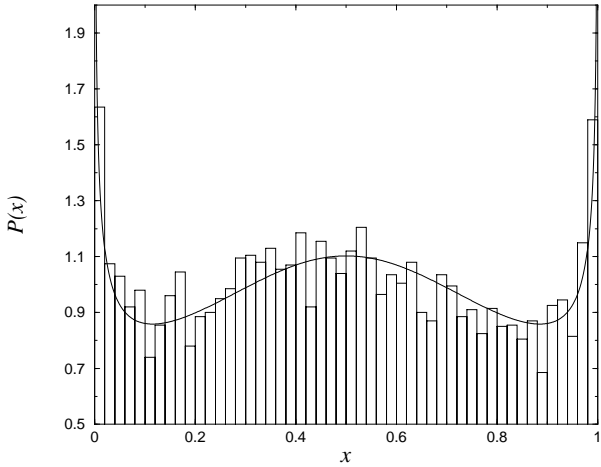}
\caption{The probability density function for the correlation function defined by the time average of $g_{10}(x)=xx_{10}$ ($B=2.5$).
}
\label{fig3}
\end{figure}

\begin{figure}[h]
\includegraphics[width=8cm]{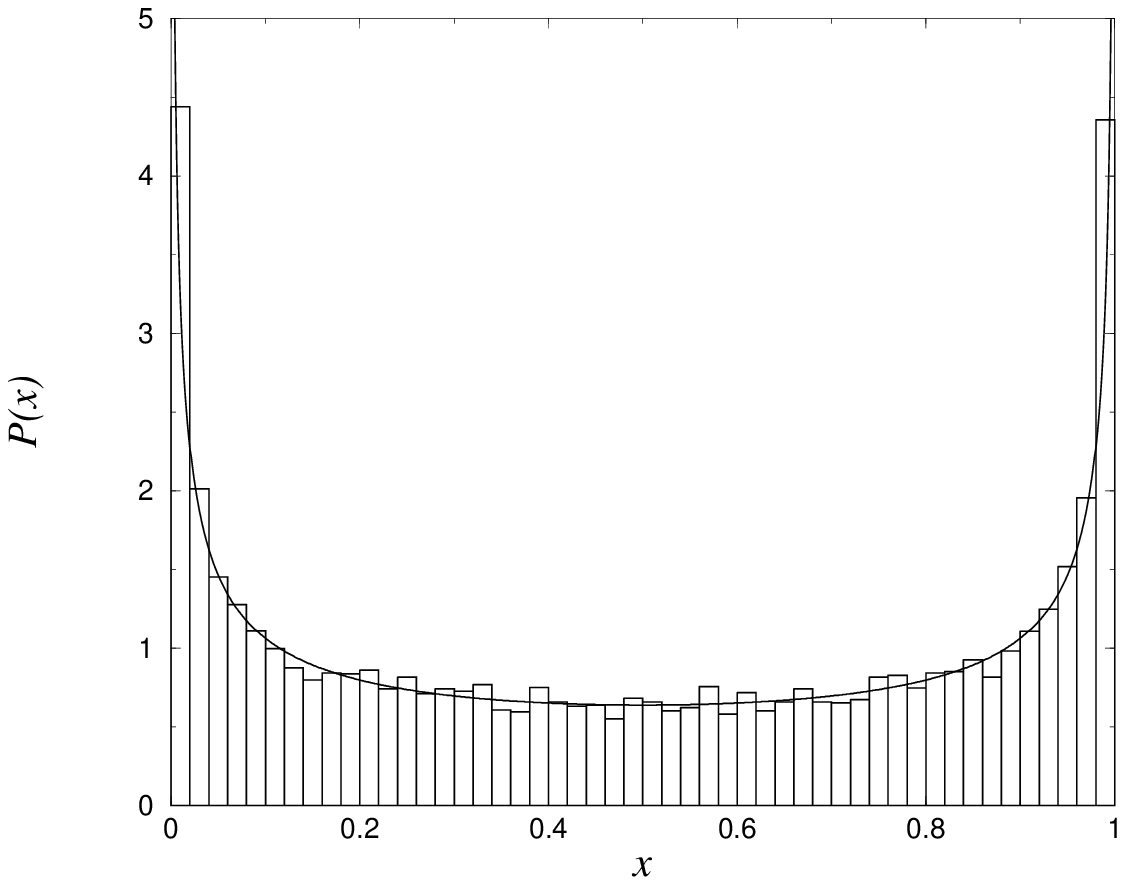}
\caption{The probability density function for the correlation function defined 
by the time average of $g_{10}(x)=xx_{10}$ ($B=3.0$).
}
\label{fig4}
\end{figure}

Figures \ref{fig3} and \ref{fig4} show that the correlation functions with 
fixed time difference $n$ are intrinsically random and these distributions 
obey the generalized arcsine distribution.

\section{Stable law}

In this section we demonstrate the distribution for 
the time average of the $L^1_{loc}(0,1)$ function with infinite mean, 
which satisfies the condition (\ref{eq:3.7}) and 
\begin{equation}
\mathop{\lim}_{\epsilon\rightarrow 0}
\frac{\int_{\epsilon}^{1-\epsilon} |g|dm}
{\int_{\epsilon}^{1-\epsilon} dm}=\infty.
\label{eq:4.1}
\end{equation}
In the case of the finite measure, the time average of the $L^1_{loc}(0,1)$ 
function with infinite mean converges to the stable distribution, as shown 
in Appendix A.\par
In what follows, we study numerically the distribution for the time average 
of the observation function
\begin{equation}
g(x)=\left\{
\begin{array}{ll}
x^{-\alpha}   &(x<\frac{1}{2})\\
(1-x)^{-\alpha}   &(x\geq \frac{1}{2}),
\end{array}
\right.
\label{eq:4.2}
\end{equation}
which is the $L^1_{loc}(0,1)$ 
function with infinite mean and $\alpha>0$.
As shown in Fig. 6, we find that the distribution for 
the scaled time average of $L^1_{loc}(0,1)$ 
function with infinite mean also converges to the stable distribution 
with index $\gamma$. In numerical simulations we calculate the time average 
of $g(x)$ for three different length of the simulation time $(n=10^4,10^5$ and 
$10^6)$ and then determine the exponent 
$\gamma$ so as to make the distributions of the scaled time average 
invariant.
Figure \ref{fig11} shows the linear relation between 
$\gamma$ and $\alpha$ clearly. Moreover,
we find that the scaling exponent $\gamma$ obeys 
the non-trivial relation to the exponent $\alpha$ and $B$ except for
the case $B< 2.5$, i.e., 
\begin{equation}
\gamma=\frac{\alpha}{B-1}+1.
\label{eq:4.3}
\end{equation}
This relation is clearly shown in Fig. \ref{fig12} except for the case 
$B<2.5$. 
\footnote{The reason that
 the relationship $(\ref{eq:4.3})$ does not hold for the case $B<2.5$
 seems to be that the observation time is enoug in numerical simulations.}
We summarize these results as following conjecture.

\begin{Con}
Let $g(x)$ be the $L^1_{loc}(0,1)$ function with infinite mean 
w.r.t. the invariant measure $m$. Further, 
\begin{equation}
x^{\alpha}g(x) =O(1), \quad x\rightarrow 0
\label{eq:4.4}
\end{equation}
\begin{equation}
(1-x)^{\alpha}g(x) =O(1), \quad x\rightarrow 1
\label{eq:4.5}
\end{equation}
 Then the scaled time average of $g(x)$ converges to the stable
 distribution $G_{\gamma}$:
\begin{equation}
\frac{1}{b_n}\sum_{k=0}^{n-1}g(T^k\cdot)
\rightarrow G_{\gamma},
\label{eq:4.6}
\end{equation}
where $b_n\propto n^{\gamma}$ and $\gamma=\frac{\alpha}{B-1}+1$.
\label{Con:3}
\end{Con}

\begin{Rem}
In the case of $B=2.0$ the scaling sequence $b_n$ had better be 
$n^{\gamma}/\log n$ rather than $n^{\gamma}$.
\end{Rem}

\begin{Rem}
The distribution for 
the scaled time average of $g(x)=x^{-\alpha}$
 also converges to the stable distribution when the 
invariant measure is finite. On the other hand, the distribution for 
the scaled 
time average of $g(x)$ is different from the stable one due to 
the generalized arcsine law for the occupation time of the interval $[1/2,1]$.
\end{Rem}

\begin{figure}[h]
\includegraphics[width=8cm]{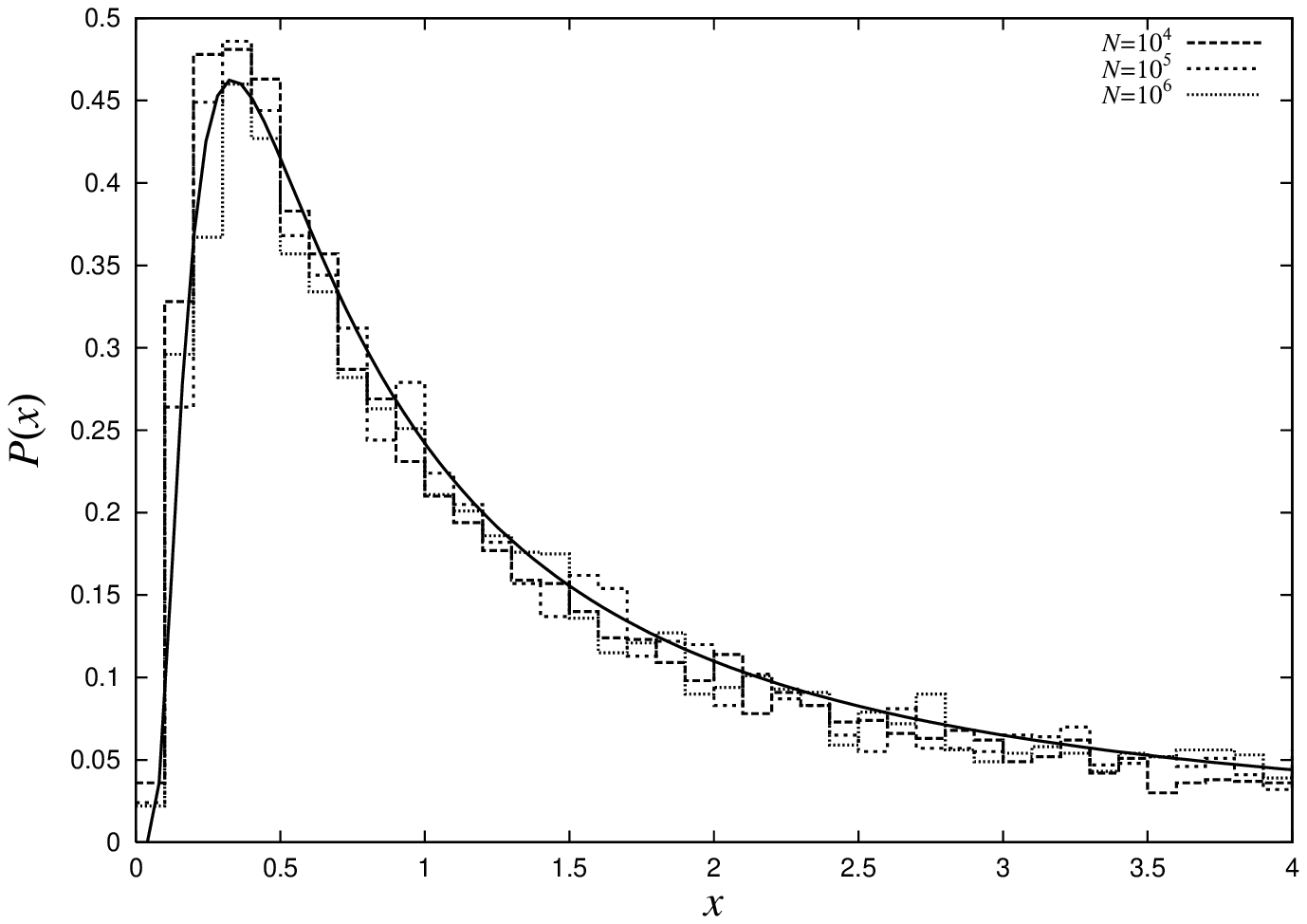}
\caption{The probability density function for the scaled time average of $g(x)$ ($B=3.0$ and $\alpha=2.0$). The fitting curve is a stable distribution with 
$\gamma=2.0$ $(P(x)=\frac{1}{\sqrt{2\pi x^3}}e^{-1/(2x)})$.
}
\label{fig10}
\end{figure}

\begin{figure}[h]
\includegraphics[width=8cm]{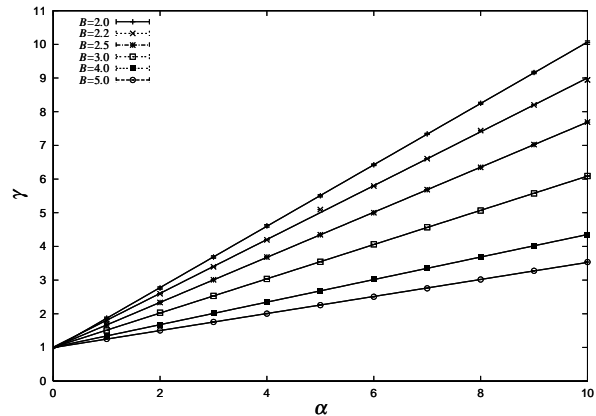}
\caption{Scaling exponent $\gamma$ vs. $\alpha$.
}
\label{fig11}
\end{figure}

\begin{figure}[h]
\includegraphics[width=8cm]{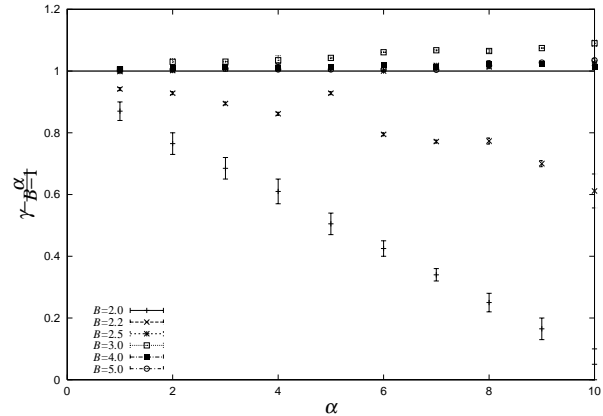}
\caption{Relationship between $\gamma$ and $\alpha$ and $B$.
}
\label{fig12}
\end{figure}

\section{Discussion}

In this paper we present the distributional limit theorems for the 
time average of the $L^1_{loc,m}(0,1)$ function with finite mean
and infinite mean
 using the modified 
Bernoulli map whose invariant measure is infinite. 
By applying this theorem to the correlation function, it is clearly shown 
that the correlation functions with fixed time difference $n$ converge 
to the generalized arcsine distribution. G. Margolin and E. Barkai analyzed
the distribution for
 the correlation function of a dichotomous random process 
 and its convergence process \cite{Marg}. 
Our results correspond to the generalized result of their work
in the way that the observation function can be extended to 
the $L^1_{loc,m}(0,1)$ function with finite mean.\footnote{ 
In \cite{Marg} the observation function is the characteristic 
function $I(x)$ with $a=1$ and $b=0$, which is the special case of our 
results.}
However, the convergence process of the 
time average of the $L^1_{loc,m}(0,1)$ with finite mean is not studied. So 
we will study the convergence process for the correlation function in 
the forthcoming paper.\par
From the viewpoint of physical observation, 
the distributional limit theorems, namely, the Mittag-Leffler distribution 
and the generalized arcsine law and the stable law, suggest that 
one can characterize the behaviour of non-stationary phenomena through 
the distribution of the time average. 
It is important to know 
what class the observation function belongs to, that is, whether 
the observation function is the $L^1_{loc,m}(0,1)$ function with 
finite mean or not. 
Because the observation function in physical systems 
is not always the $L^1(m)$ function.
Actually, we show that
the correlation function is a typical example of the 
$L^1_{loc,m}(0,1)$ 
function with finite mean. 
This means that the correlation function, or 
the statistical quantities based on the time average in the 
intermittent phenomena, is intrinsically random. Universal distributions 
for the time average of various observation functions are shown in Table 
\ref{tab:1}.
\par
In the concept of the \textquotedblleft ergodicity\textquotedblright
 proposed by L. Boltzmann, 
the ergodicity, i.e., the time average equals to the space average, 
guarantees the existence of the equilibrium state in dynamical 
systems. In the non-equilibrium steady state, Sinai-Ruelle-Bowen
 measure is considered 
to describe the non-equilibrium steady state. However, there are no concepts of the \textquotedblleft ergodicity\textquotedblright in the non-equilibrium
 state, that is, the measure of 
the non-equilibrium non-stationary state is not at all 
elucidated on the basis of the time average of the dynamical systems.
\footnote{From the aspect of the ensemble average,
the approach to equilibrium of the observables
is clearly shown using infinite measure systems in \cite{Ak1,Ak2,Tasaki}. }
We hope that randomness of the time average in infinite measure 
systems will give us the motive argument toward the ergodic 
problem in the non-equilibrium state. 
There is a possibility that 
Mittag-Leffler distribution or
the generalized arcsine distribution or the stable distirbution 
could become one of the measures 
 characterizing the non-equilibrium state. 
Actually, these distribution universally appear
in diffusion and its generalizations. Moreover, the generalized arcsine 
law has drawn much attention in disordered systems \cite{Bar,Bur}.


%
%

\begin{table*}
\caption{Universal distributions of the time average of the observation function $g(x)$.}
\label{tab:1}       
\begin{tabular}{lll}
\hline\noalign{\smallskip}
Invariant measure & $g(x)$ & Distribution  \\
\noalign{\smallskip}\hline\noalign{\smallskip}
Finite ($B<2$)& $L^1(m)$ & {\it Delta} \\
Finite ($B<2$)& $L_{loc}^1(m)$ with infinite mean &  {\it Stable}\\
Infinite ($B\geq 2$)& $L^1(m)$ & {\it Mittag-Leffler} \\
Infinite ($B\geq 2$)& $L_{loc,m}^1$ with finite mean & {\it Generalized arcsine} \\
Infinite ($B\geq 2$)& $L_{loc,m}^1$ with infinite mean & {\it Stable} \\
\noalign{\smallskip}\hline
\end{tabular}
\end{table*}

\begin{acknowledgements}
T.A. would like to thank Y. Aizawa and T. Inoue for fruitful discussions.
T.A. is supported by a grant to The 21st Century COE Program (Physics of 
Self-Organization Systems) at Waseda University from the Ministry of 
Education, Culture, Sports, Science and Technology, Japan.
\end{acknowledgements}

\appendix
\section{The distribution for the time average of the non-$L^1(m)$ function 
in the case of the finite measure}

In the case of the finite measure $(B<2)$, the invariant density can be 
written as 
\begin{equation}
\rho (x)=\frac{2-B}{2}\{x^{1-B}+(1-x)^{1-B}\}.
\label{eq:a.1}
\end{equation}
Birkhoff's ergodic theorem \cite{Bir} tells us that 
the p.d.f. of the sequence $\{Tx, T^2x, \cdots, T^nx\}$ 
obeys the invariant density as $n\rightarrow \infty$.
Let ${\bf X}_n$ be random variables with p.d.f. $(\ref{eq:a.1})$ 
and $g(x)=x^{-\alpha}~(\alpha\geq 2-B)$. 
The distribution of ${\bf Y}_n=g({\bf X}_n)$ is given by 
\begin{eqnarray}
F(x)=\Pr({\bf Y}< x)
&=&\Pr({\bf X}>x^{-\frac{1}{\alpha}})\nonumber\\
&=&1-x^{-\frac{2-B}{\alpha}}.
\label{eq:a.2}
\end{eqnarray}
The general central limit theorem \cite{Feller} says that 
the random variable 
$
({\bf Y}_1+\cdots +{\bf Y}_n)/n^{\alpha/(2-B)}$
has the stable distribution.
Therefore 
\begin{equation}
\frac{1}{n^{\frac{\alpha}{2-B}}}\sum_{k=0}^{n-1}
g(T^k\cdot)\rightarrow G_{(2-B)/\alpha}~{\rm as}~n\rightarrow \infty,
\label{eq:a.3}
\end{equation}
where $G_{(2-B)/\alpha}$ is the stable distribution with index 
$(2-B)/\alpha$.



\end{document}